\documentclass[pss]{wiley2sp} 
\usepackage{amsmath}
\usepackage{hyperref}
\hypersetup{colorlinks=true,citecolor=black,linkcolor=black,urlcolor=black}
\newcommand{\etal}{\textit{et al}.\ }
\begin{document}

\title{Transporting and manipulating single electrons in surface-acoustic-wave minima}

\titlerunning{SAW-driven quantum dots}

\author{%
  Christopher J.\ B.\ Ford\textsuperscript{\Ast
  	}
  }

\authorrunning{Ford}

\mail{e-mail
  \textsf{cjbf@cam.ac.uk}, Phone: +44 1223 337486, Fax: +44 1223 337271}

\institute{%
  Cavendish Laboratory, J J Thomson Avenue, Cambridge CB3 0HE, United Kingdom\\
  }

\received{XXXX, revised XXXX, accepted XXXX} 
\published{as Phys. Status Solidi B, 1600658 (2017) / DOI 10.1002/pssb.201600658} 


\abstract{%
%
%
%
\abstcol{%
  A surface acoustic wave (SAW) can produce a moving potential wave that can trap and drag electrons along with it. We review work on using a SAW to create moving quantum dots containing single electrons, with the aims of developing a current standard, emitting single photons, transferring single electrons between static quantum dots, and investigating non-adiabatic effects.}{%
  }}

%
%
\titlefigure[height=3.1cm,keepaspectratio]{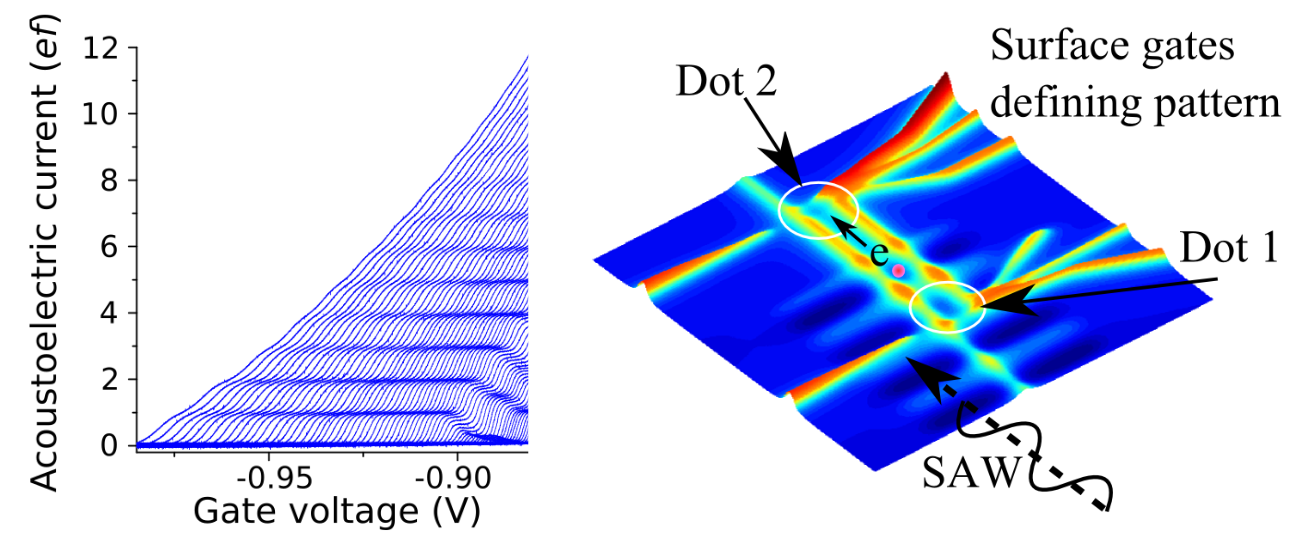}
\titlefigurecaption{%
  }


\maketitle   

\section{Introduction}In a piezoelectric material, a surface acoustic wave (SAW), which is a strain wave moving along the surface, is accompanied by a potential wave. Charge can be trapped in the potential minima and dragged along, even being given energy from the sound wave to drive it up a slope in the external potential. Over the last twenty years, this phenomenon has allowed us to manipulate single electrons, by confining them in one-dimensional (1D) channels using external potentials and isolating them from each other along the wire with the SAW potential. The stream of single electrons produces a current $nef$, where $n$ is the number of electrons in each SAW minimum, $e$ the electronic charge and $f$ the SAW frequency. It was long hoped that this quantised `acoustoelectric' current could be accurate enough to become a new standard of current, since the frequency (in the GHz range) is high enough to drive a current that can be measured sufficiently accurately for such a standard. However, current plateaux have never been flatter than about 1 part in $10^4$, and we have turned our attention to other uses of these dynamic quantum dots. Useful reviews of SAW-driven current and other techniques for producing quantised currents were written by Flensberg \etal in 1999 \cite{FlensbergReview1999} and Kaestner in 2015 \cite{KaestnerReview2015}.

A potential minimum moving past a static (gate-defined) quantum dot can pull an electron out of the dot, and transfer it back and forth between such dots. Each electron in a stream of single electrons feels a rapidly varying potential around it as it passes into a channel with a different width. This can cause a non-adiabatic excitation from the ground state into a combination of ground and excited states, and the resulting coherent oscillations can be detected by small variations in the amount of tunnelling into an adjacent channel. In the longer term it should be possible for each electron to combine with a hole to give out a single photon, and the spin of each electron could be manipulated by static magnets and tunnel barriers while passing along the channel, as a form of quantum processor or quantum repeater. 

The following sections will describe some of the work on each of these aspects of the transport and manipulation of single electrons in moving quantum dots. Much of this work was carried out at the Cavendish Laboratory in Cambridge.

\begin{figure}[htb]%
	\includegraphics*[width=\linewidth]{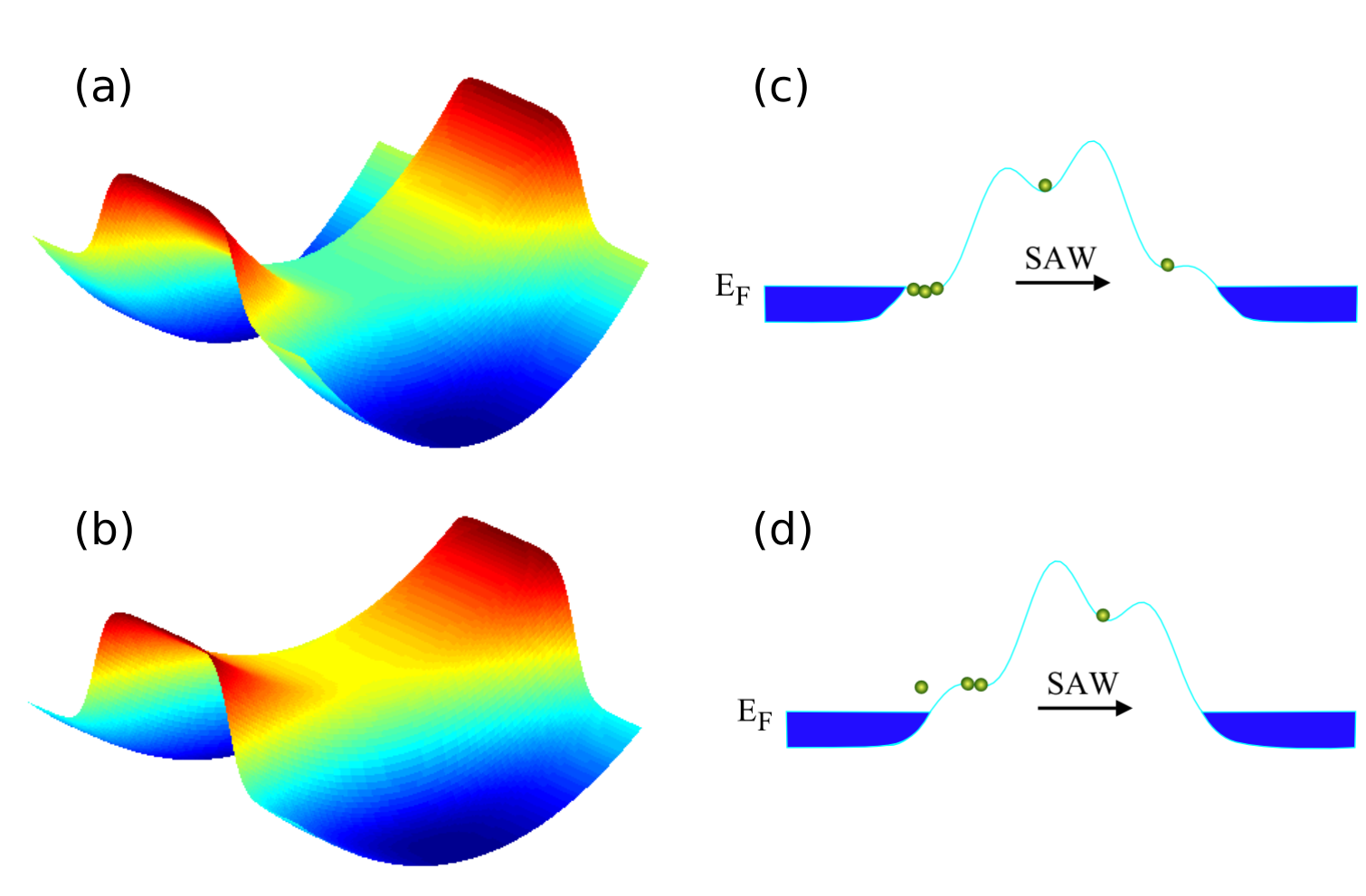}
	\caption{%
		(a) Schematic view of the device structure, showing the 2D layer of electrons (2DEG) just below the surface, which is depleted in a pattern defined by negative voltages on split gates on the surface, forming a 1D channel that can be depleted completely if required. (b) Schematic intensity plot of the potential around an open channel. The potential is low (blue) in the 2DEG leads, and high (red) under the gates depleting the electrons from underneath the gates to define the channel. The potential is higher in the channel (green) than in the leads but not enough to bring it above the Fermi energy. (c) As in (b), but with a more negative gate voltage, so that even electrons in the channel are depleted. (d) Section through the potential landscape along a depleted channel. The SAW is moving from left to right and its potential minimum is dragging a few electrons (circles) out of the 2DEG. (e) As in (d), at a slightly later time when the dot containing the electrons is being squeezed between the hill and the SAW peak to its left. Electrons leave one by one until the steepest part of the potential is reached, after which the dot starts to get larger again and any surviving electrons are carried over the hill to the 2DEG on the right, producing a current.}
	\label{potentialdiagrams}
\end{figure}

\begin{figure}[t]%
	\includegraphics*[width=\linewidth]{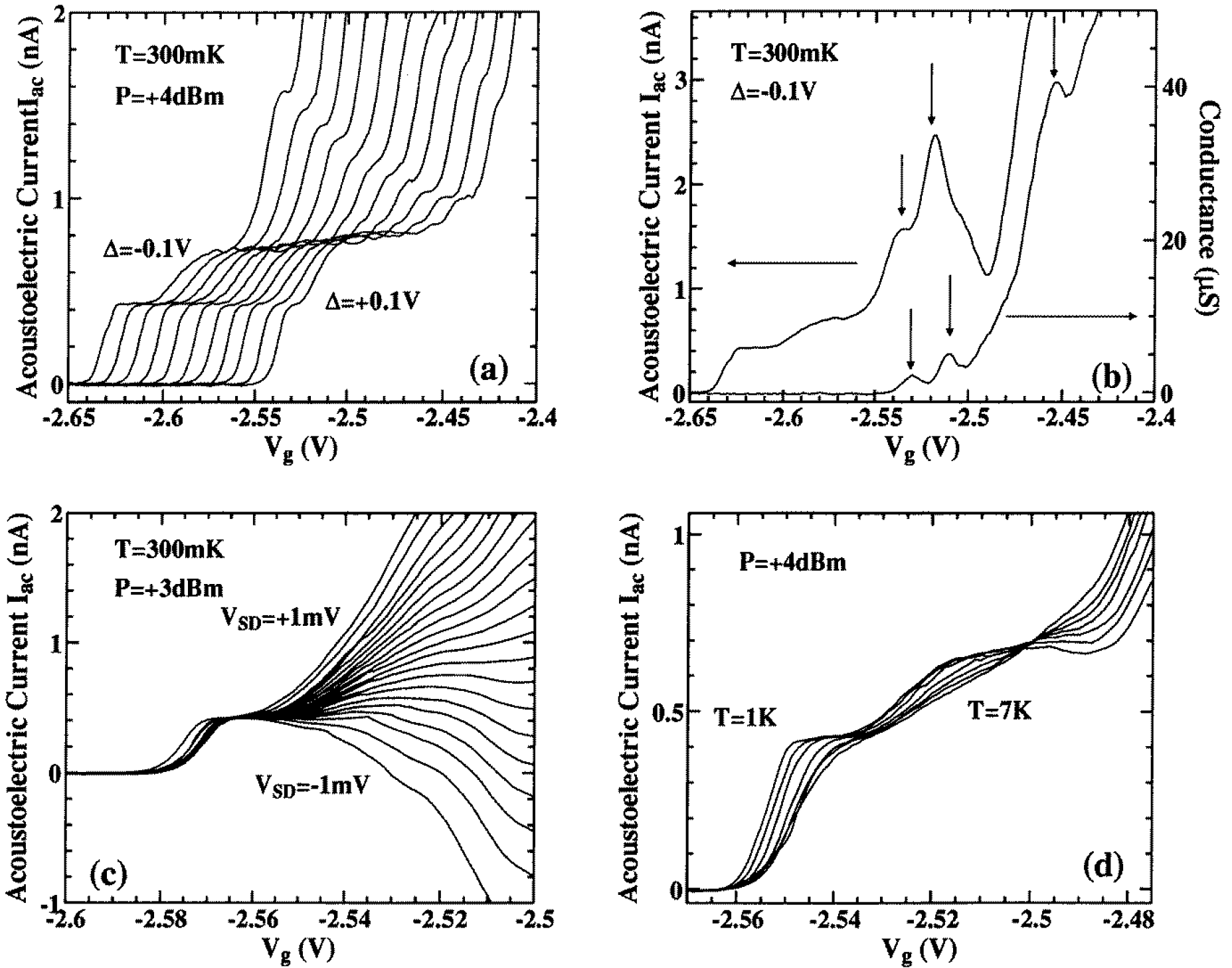}
	\caption{%
		(a) The acoustoelectric current versus split-gate voltage $V_{\rm g}$ with voltages on the gate arms which differ by $\Delta$, which moves the channel sideways. (b) Comparison of the acoustoelectric current and the conductance. (c) The acoustoelectric current \textit{versus} split-gate voltage at different source-drain biases $V_{\rm SD}$. (d) The acoustoelectric current versus split-gate voltage at different temperatures $T$. $P$ is the applied SAW power. \href{http://dx.doi.org/10.1088/0953-8984/8/38/001}{Reproduced with permission from Ref.\ \cite{Shilton1996-484}, copyright 1996 by IOP Publishing, all rights reserved}.}
	\label{figShilton1996-484-f5}
\end{figure}

\section{Quantised acoustoelectric current}
The idea that a SAW could drive a current through a conductor was proposed by Parmenter in 1953 \cite{Parmenter1953}, by considering that the small proportion of electrons whose velocity matched the SAW velocity would be carried along in the potential minima arising from the deformation potential of the crystal distorted by the acoustic wave. This was shown experimentally four years later by Weinreich and White \cite{Weinreich1957} in a piece of \textit{n}-type germanium.

\sloppy The first attempt to pass a SAW along a two-dimensional (2D) layer of electrons near the surface of a crystal was by Wixforth \etal \cite{Wixforth1986}, using the 2D electron gas (2DEG) formed in the potential well at the interface between layers of GaAs and AlGaAs. This type of heterostructure forms the basis for all the work described here. The electrons lie close to the surface so they all experience the potential arising from the SAW. There is very little scattering of electrons, as the ionised donors are spatially separated from the 2DEG, and the mean free path can be considerably greater than the SAW wavelength. The SAW potential can therefore give considerable momentum and energy to the electrons before they lose it to the lattice through scattering off impurities, particularly at low densities when the conductivity is close to a characteristic value related to the SAW velocity. This was first detected as increased attenuation of the SAW \cite{Wixforth1986} and later as an acoustoelectric (SAW-driven) current \cite{Esslinger1992}, both at high magnetic fields. At low fields, where the electrons strongly screen the SAW, Shilton \etal found the acoustoelectric current to be enhanced when the SAW wavelength matched the cyclotron diameter, trapping electrons in the SAW potential minima \cite{Shilton1995-1879}. 

Shilton \etal next patterned the surface with a split pair of gates, that were used to deplete electrons underneath them to leave a narrow, 1D, channel along the SAW propagation direction \cite{Shilton1996-1925} (see Fig.\ \ref{potentialdiagrams}(a)). This exhibited strong oscillations in the acoustoelectric current as the number of 1D subbands in the channel was reduced by making the gates more negative to squeeze the channel. It was a natural next step to deplete this channel completely, in the hope that the SAW potential would drag electrons from the 2DEG at the entrance to the channel, and drag them along the channel. For a channel longer than the SAW wavelength, the SAW potential would provide a barrier behind and in front of the electrons, confining them in quantum dots---regions small enough to have discrete energy levels and a large Coulomb charging energy (see Fig.\ \ref{potentialdiagrams}(d) and (e)). Adding another electron would cost this charging energy and so there should be a range of gate voltages over which it was energetically favourable only to have a certain number of electrons in the dot. This is exactly what was eventually observed \cite{Shilton1996-484}. A plateau appeared in the acoustoelectric current at about 450\,pA, corresponding to $ef$, for SAW frequency $f=2.7$\,GHz, independent of the applied bias or SAW amplitude (see Fig. \ref{figShilton1996-484-f5}). 

Maps of the potential around the channel are shown in fig.~\ref{potentialdiagrams} when open (b) and depleted (c), producing an empty 1D channel within which there will be no screening of the SAW (except by the side gates). The schematic potential along the channel is shown in (d) and (e) at consecutive time steps, showing how electrons are dragged out of the 2DEG lead at the left as the screening by the 2DEG decreases its edge, and then driven up the potential hill created by the side gates. The falling SAW potential is added to the rising channel potential, and so the resulting quantum dot has its minimum depth where the hill is steepest. At this point, any electrons that are not ejected stay inside until the SAW becomes screened again at the 2DEG on the right. This flow of electrons constitutes the acoustoelectric current. If the same number $n$ of electrons remains in each consecutive SAW minimum, the current will be quantised at $nef$. However, since the observed quantisation is poor, there must be error mechanisms, such as potential instabilities due to fluctuating occupancy of traps, insufficient time for excess electrons to tunnel out, or a chance of captured electrons tunnelling out. The accuracy and error mechanisms will be discussed later.

The following year, in 1996, Julian Shilton, with the rest of our team in Cambridge, led by Mike Pepper and Valery Talyanskii, improved the quantisation accuracy and observed at least four moderately flat plateaux \cite{Talyanskii1997-1963}. Detailed measurements across the first plateau showed a great decrease in noise on the plateau, and a flatness of about 0.3\%. The noise was explained as potential instability due to one or more electrons tunnelling (`switching') randomly back and forth between trapped states in or near the channel, which effectively changes the gate voltage randomly between two or more values, producing a `random telegraph signal' (RTS). On a plateau such a shift in gate voltage has less effect, but near the ends of a plateau it averages plateau and riser, and hence causes the quantisation to be poor and the plateau length to be shorter than it should be. (Methods of reducing such switching noise have since been devised, by cooling samples down from room temperature with a positive gate bias applied \cite{PioroLadriere2005}. RTS noise in SAW devices has been studied by Song \etal \cite{SongRTS2012}, and the authors concluded that increasing the SAW power greatly reduced the likelihood of an RTS as it made the quantum do potential well much deeper. Unfortunately, this did not result in improved quantisation.) 

\begin{figure}[t]%
	\includegraphics*[width=\linewidth]{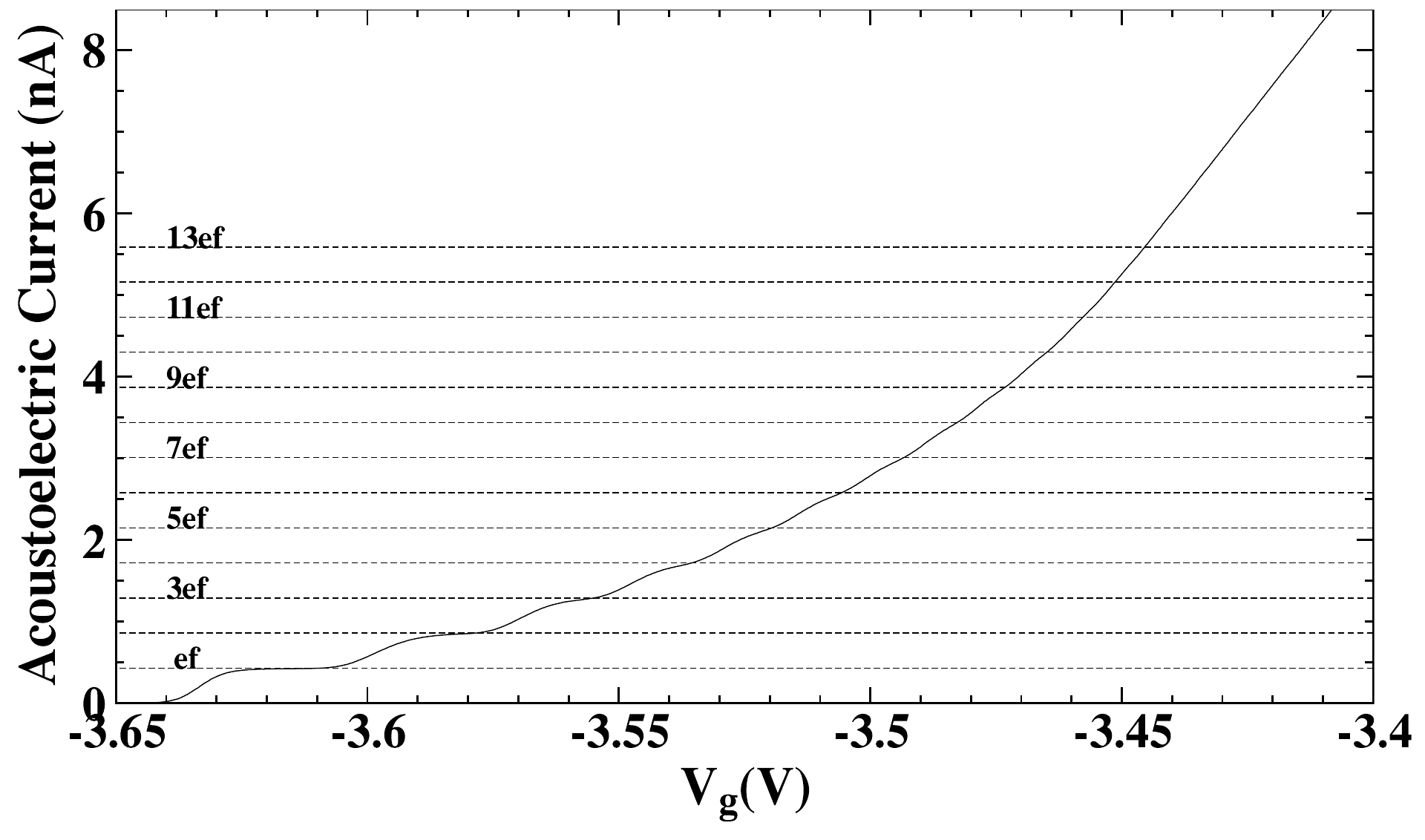}
	\caption{%
		An example of the quantised acoustoelectric current measured at a temperature of 1.2\,K after optimisation by tuning the applied microwave frequency and power. From \cite{CunninghamThesis2000}, copyright 2000 by J.\,E.\,Cunningham.}
	\label{Cunninghamplateaux}
\end{figure}

\begin{figure}[t]%
	\includegraphics*[width=\linewidth]{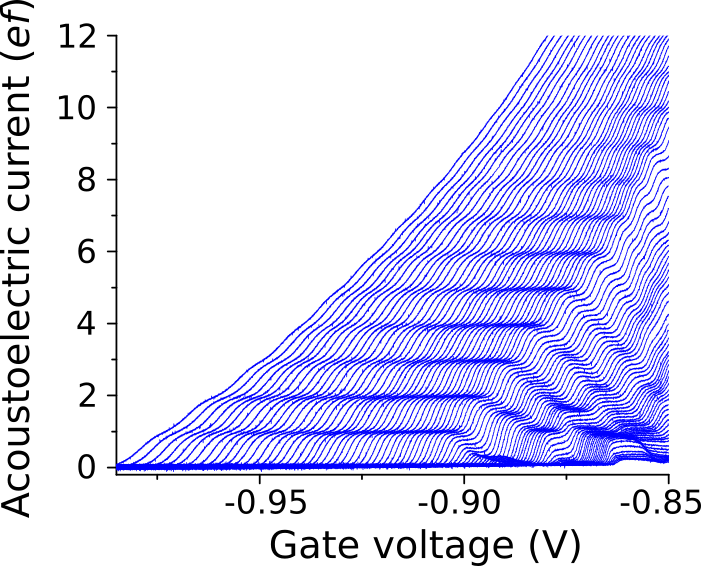}
	\caption{%
		An example of acoustoelectric current plateaux. The power applied to the transducer is varied from 13\,dBm (left) to $-5$\,dBm (right) in steps of 0.2\,dBm. From \cite{AstleyThesis2008}, copyright 2008 by M.\,J.\,Astley.}
	\label{Astley_Michael_PhD2.5}
\end{figure}

Over the following years much effort was put into trying to increase the flatness of the plateaux. Some examples are shown in Figures \ref{Cunninghamplateaux} and \ref{Astley_Michael_PhD2.5}. Unfortunately, no plateaux flatter than 1 in $10^4$ have been observed and attention turned to other applications of such moving quantum dots. Cunningham \etal investigated wires defined by shallow etched trenches on either side rather than by gates \cite{Cunningham2000-2239}, finding fewer problems with switching noise. Such channels have stronger lateral confinement than usually obtained with split gates, and this may have made it easier to get the best precision on the plateau down to 60\,ppm. This is still some way off from the $<1$\,ppm required for a current standard. At the UK standards lab, the National Physical Laboratory (NPL), Janssen and Hartland set up a cryogenic current comparator to measure the current with better than 50\,ppm accuracy, and tested SAW devices \cite{Janssen2000}, finding as usual that RTS noise was a major limitation.
A proposal to `top up' empty dots with electrons as they pass by a side tunnel barrier \cite{Talyanskii1998-375} in order to increase the pump accuracy has not yet been tried in earnest.

A SAW is generated in a piezoelectric material using an interdigitated transducer consisting of two sets of metal fingers, interleaved so that a radio-frequency signal applied to one set, with the other set grounded, produces an oscillating electric field between all the fingers. This produces a strain wave coupled to a potential wave. The transducer has a high quality factor and therefore a narrow resonance. A convenient periodicity of the fingers is $\lambda=1\,\mu$m, so the resonance occurs at $f\approx 2.7$\,GHz, when $v_{\rm SAW}=f\lambda$, where $v_{\rm SAW}$ is the SAW velocity ($\sim 2700$\,m\,s$^{-1}$). In each sample it has always been possible to optimise the quantisation accuracy by varying the SAW frequency slightly across the resonance. This was shown to be periodic with a period corresponding to beating between the main SAW and the weaker wave reflected from a second (spare) identical transducer on the other side of the device. In addition to a large travelling wave, the combination produces a low-amplitude standing wave, whose antinodes can be moved through the channel by changing the frequency slightly. The variation of the quantisation with frequency therefore indicates that there is a moment as the electron moves up the potential hill where an increase or decrease in SAW potential at the antinode can help to expel unwanted electrons and/or to keep the rest. In order to tune this effect, Cunningham \etal \cite{Cunningham1999-2060} applied the same radio-frequency signal to the second transducer, with variable amplitude and phase. For a particular amplitude, the $ef$ plateau could easily be destroyed at most phase shifts, but at a certain phase shift the residual slope on the plateau was smaller by a factor of four than without any intentional counter-propagating SAW. Utko \etal \cite{Utko2007} later studied the effect of such a pair of SAWs in detail in an etched semicircular channel, seeing many plateaux but also finding an extra pumping mechanism probably related to an unintentional quantum dot formed in the short central part of their channel. Counter-propagating SAWs were investigated further recently by He \etal \cite{HeSAWinB2014} in conjunction with a magnetic field (see below). There is scope for further work, perhaps with multiple split gates in series, or pulsed gates, to gain more control of the pick-up process instead of using the counter-propagating SAW.

Theoretical work on quantised SAW pumping has been quite limited. In 1998, an analytical quantum-mechanical solution was derived by A\v{\i}zin, Gumbs and Pepper \cite{Aizin1998}, for a single electron in a 1D potential, showed that the probability of the electron tunnelling out rapidly decreased to zero, leading to an $ef$ plateau. The paper also showed that the SAW potential under a surface gate was strongly screened but retained its shape. Shortly after this, Flensberg \etal \cite{Flensberg1999} modelled the capture process quantum mechanically, taking account of the fact that the motion is non-adiabatic and so the probability of tunnelling out through the barrier behind the moving dot decreases exponentially with time. This exponential variation introduces a characteristic time and hence an energy uncertainty. There is then an effective temperature $T_{\rm eff}$, where $k_{\rm B}T_{\rm eff}$ is the larger of this uncertainty and the thermal energy ($k_{\rm B}$ is the Boltzmann constant). The slope $S$ of the current plateau is given by $S\approx (2E_{\rm c}T_{\rm eff})e^{-E_{\rm c}T_{\rm eff}}$, where $E_{\rm c}$ is the charging energy of the moving dot. As a result, both the slopes of plateaux and the widths of the transition regions between the plateaux saturate at low temperature at the values determined by the characteristic energy scale for non-adiabatic corrections.

Robinson and Barnes then modelled the electron pick-up process using a \textit{classical} model of point charges interacting in a moving, squeezed dot with a realistic potential \cite{Robinson2001-2241}. The dot could hold about 20 electrons at the moment when it became separated from the 2D lead. As it was pushed up the potential slope of the depleted channel by the SAW peak, it became smaller, and this compression did work on the electrons, heating them so that after a few picoseconds there was enough excess total energy in the dot, including the potential energy from the repulsive Coulomb forces between all the electrons, to expel one electron over the energy barrier caused by the SAW peak (as in fig.~\ref{potentialdiagrams}e). The interaction energy then dropped and the number of electrons in the dot was stable for a few more picoseconds until the process repeated as the dot shrunk. At the steepest part of the hill, the dot was at its smallest, and any electrons that stayed in the dot past this point were then carried over the hill to the other lead. There was thus a critical position in the channel where the final number of electrons $n$ was determined. In a few picoseconds the ($n$+1)th electron had to leave and then there was a further similar amount of time over which the $n$th electron had a chance of leaving. As mentioned earlier, a counter-propagating SAW provides a crude way of changing one or other of these two probabilities \cite{Cunningham1999-2060}. Even though this model is classical, it is very instructive and an equivalent quantum-mechanical model (which is hard since interactions between electrons in the dot must be taken into account) would probably give very similar results, but tunnelling would increase the probabilities of each electron leaving.

As with the single-particle quantum-mechanical modelling of Flensberg \etal \cite{Flensberg1999}, this classical, many-particle modelling showed that there was a characteristic temperature in the dot $T_{\rm eff}$: as each excess electron was forced out, for $T<T_{\rm eff}$, the remaining electrons fell into the potential well left behind, and half of the energy gained became kinetic energy through equipartition, causing heating. Above $T_{\rm eff}$, evaporative cooling dominated instead. Thus the electron temperature always moves towards $T_{\rm eff}$. $T_{\rm eff}$ was estimated to be about 1.7\,K for typical experimental parameters. These classical and quantum temperature scales do not seem to be related so they may both be relevant in real devices. A higher temperature means that the probability of a trapped electron leaving the dot is higher, so some dots are empty, and this results in sloping plateaux, rising slowly to $nef$. This also means that the sample temperature must be kept low (below 4\,K) and heating from the power applied to drive the SAW must be limited \cite{Fletcher2002}. There is no experimental evidence for these temperature scales, but quantised current plateaux are seen at 4\,K and below, but do not improve significantly at temperatures lower than $\sim 1$\,K, which is consistent with the theories. In 2009, Guo \etal \cite{GuoModelling2009} modelled the SAW potential using a realistic 3D self-consistent Poisson-Schr\"{o}dinger solution. They found that the current quantisation should be best for a channel a little longer than the SAW wavelength. The details of the SAW potential itself have been modelled by Rahman \etal in 3D using finite-element methods \cite{Rahman2006-2844,Rahman2007-2871}.

When a magnetic field is applied perpendicular to the surface, it is often found that the acoustoelectric current increases dramatically for fields above $\sim 0.2$\,T \cite{Cunningham2000-2239}, which is also roughly the field at which quantum-Hall edge states usually start to become well-defined \cite{Ford1994-1835}. It was not clear why the field suppresses the \textit{loss} of excess electrons as they are dragged up the hill by the SAW (as in Fig.\ \ref{potentialdiagrams}(e)). Very recently, He \etal \cite{HeSAWinB2014} have investigated this in detail, as a function of the phase of an additional SAW sent from the other transducer (as described above \cite{Cunningham1999-2060}). They showed that excited states of the quantum dot become visible as features in the differentiated current at finite magnetic fields. The explanation the authors offer is that the SAW quantum dot is elongated along the propagation direction, and so the magnetic field can affect the longitudinal motion even at fields below 0.5\,T. Non-adiabatic perturbations of the potential can easily excite electrons into the higher states, which may keep them away from the rear barrier and hence prevent them tunnelling back into the source lead, so that they get stuck in the dot. The perturbations can come from the movement of the SAW past gates \cite{Kataoka2009-3029}, the magnetic field, or the SAW reflected from the other transducer.

To assess the number of SAW-driven dots that either have an extra electron or a missing one, Robinson \etal measured the shot noise for frequencies up to 2\,kHz \cite{Robinson2002-2358}. Away from the acoustoelectric current plateaux, the noise was larger than expected for shot noise, owing to random-telegraph switching as mentioned above. However, on the $ef$ plateau, the noise became immeasurably small, implying that fewer than 10\% of the dots carried zero or two electrons. Robinson \etal later measured noise at around 1.7\,MHz using a bespoke cryogenic amplifier \cite{Robinson2004-2561} and deduced the separate probabilities of zero and two electrons in each dot as a function of position on the $ef$ plateau \cite{Robinson2005-2716}, and it appeared that the two probabilities might be correlated (surprisingly), as the sum stayed constant along the plateau. The plateaux were not especially good so the actual probabilities were not representative of those to be expected in the best devices. The noise was predominantly random telegraph switching, which followed the gradient of the current very well, but after subtracting this contribution, the shot noise related to errors in the occupation of each dot could be seen to reach a minimum on the plateau, as expected.

Crook \etal used `gated-charge force microscopy' (detecting the change in force on a probe tip as the SAW was turned off and on) \cite{Crook2008-3150} to examine a channel during SAW pumping and found that there were about four bumps in the potential along a short (1\,$\mu$m) channel, indicating that disorder in the depleted, and therefore unscreened, channel could be a serious problem. They later used scanning-gate microscopy (detecting the effect on the acoustoelectric current as the voltage on a scanned tip was varied) to show that the current through the channel was actually controlled by the steep potential slope at the entrance to the channel and was therefore fairly insensitive to the disorder in the channel itself \cite{Crook2010-3318}.

\sloppy Crosstalk between the SAW and the electromagnetic (EM) wave emanating from the SAW transducer can be a serious problem, which has to be minimised by the design of the sample holder. Another form of crosstalk is reflection of the SAW itself as a counter-propagating SAW from the spare transducer at the other end of the sample. This can sometimes be turned to one's advantage as described above \cite{Cunningham1999-2060}, but otherwise it just complicates matters and usually makes plateaux worse. Instead of a continuous wave, Kataoka \etal \cite{Kataoka2006-2858} used a SAW pulse, short enough that the EM wave from the end of the pulse would have passed the channel before the acoustic wave arrived (travelling $10^5$ times more slowly than light). For 1\,mm distance from transducer to channel, the pulse must be $<300$\,ns long. 600\,ns later the SAW pulse would return from the other transducer (2\,mm round-trip), with reduced amplitude. Further reflections from each transducer occur but are progressively smaller, though Astley \etal showed using pulsed SAWs that multiple reflections can be seen for over $400\,\mu$s in some cases, and that transducers can have power reflection coefficients as high as 0.97 \cite{Astley2006-2873}.

By using a long delay time between pulses, or one that is not an integer multiple of the travel time, the effect of the reflected SAWs can be almost eliminated, though in doing so the current becomes less by the ratio of the pulse and delay times. Also, the quantisation of the current is limited by the rise and fall times of the SAW pulse, which makes the occupation of the first and last few SAW minima uncertain as the amplitude is not optimal---a SAW with wavelength only 1000 times smaller than the distance from the transducer to channel can never be quantised to better than 1 part in about 500, even if the pulse turns off and on in the space of only one wavelength \cite{Kataoka2006-2858}. With today's high-frequency gates, it could instead be possible to synchronise a gate turning on or off between SAW cycles to let through a precise number of full-size SAW minima \cite{Hermelin2013,TalyanskiiFidelity2008}. Another measurement problem is rectification of the radio-frequency signal (applied to the SAW transducers). Giblin \etal have studied how this occurred in similar devices, taking into account the non-linearity of the dependence of current on voltage, and the capacitive coupling between gates and leads \cite{GiblinRect2013} and Howe \etal have developed a model of rectification effects that may be useful for improving device layouts to minimise these effects \cite{Howe2017}.

More complex gate geometries have also been tried. Ebbecke \etal used three pairs of split gates in series to vary the entrance and exit shapes of the channel \cite{EbbeckeGeom2002}. As expected, there was no dependence of the quantised current plateaux on the exit shape, though the SAW power needed (but not the rather poor quantisation) did depend on the shape of the entrance. As the SAW potential minimum drags electrons up the potential slope of the depleted channel, the minimum becomes smallest at the point of steepest slope, so this is where the number of electrons pumped along the channel is determined. Kataoka \etal confirmed this using a device with two pairs of split gates in series, so that the steepest slope could either occur at the front of the first or second pairs \cite{Kataoka2006-2896}. Ebbecke \etal also studied the limitations to quantisation caused by temperature and entrance shape \cite{EbbeckeLimits2003}.
To increase the total current on a plateau, either the frequency can be increased (which generates more crosstalk and other experimental difficulties \cite{EbbeckeFreq2000}), or a number of channels can be placed in parallel. Ebbecke \etal used two etched channels to double the current, but only with channels showing poor quantisation \cite{EbbeckeDouble2000}.

It is not just the popular GaAs/AlGaAs materials system that has been shown to exhibit quantised current. Astley \etal \cite{Astley2008-3038} showed $ef$ and $2ef$ current plateaux in a device that included a layer of In$_{0.1}$Ga$_{0.9}$As in the
quantum well to provide a relatively high electron $g$-factor, with an eye to future spintronic and quantum-information applications. SAW-driven quantised current was sought in carbon nanotubes by Leek \etal \cite{Leek2005-2739}, and Buitelaar \etal \cite{Buitelaar2006-2864,Buitelaar2008-3073}, but while pumping of electrons and holes was demonstrated, an $ef$ feature was only observed in one of the three samples where the contacts to the nanotube were good enough to ensure a supply of electrons in each potential minimum \cite{Buitelaar2006-2864}. W\"{u}rste \etal also found a single $ef$ plateau in their carbon nanotube sample \cite{Wuerste2007}.

In recent years, the group at Sichuan University has studied various effects in SAW-driven dots: 
He \etal \cite{HeCorrChannels2010} looked at a large ($30\,\mu$m) region with a depleted channel on either side. While they claimed they could observe Coulomb interactions in the dot as it moved through the open region between entrance and exit channels, there is also the likelihood of unmatched pumping through the two channels. This would cause charge to build up in the central region, until either pumping in becomes less efficient, or pumping out becomes easier. The group discussed this charging effect in more detail in a similar device in 2013 \cite{ChenJAP2013}, also finding an accidental `fractional' plateau. This fairly flat plateau, occurring at a fraction of the quantised value $ef$, was attributed to an unintentional static dot near the entrance to the depleted channel, as follows: on the plateau, the SAW picks up a quantised number of electrons in each cycle, but there is a finite, gate-voltage-independent, probability of an electron tunnelling back from the moving dot into the static dot and hence into the source. This fractional plateau was investigated in more detail by He \etal, who found that changing the phase of a counter-propagating SAW can move the current from one quantised plateau to the next \cite{HeSAWinB2014}. Very recently, Chen and Song confirmed that electrons were being lost from the moving dot via a static impurity dot, and calculated tunnelling-rate equations for this process \cite{ChenEPL2016}. The systematic drop in all the plateau values seen by Astley \etal \cite{AstleyThesis2008}, see the right-hand side of Fig.\ \ref{Astley_Michael_PhD2.5}, may have a similar explanation, though it is possible that there is another error mechanism that does not depend strongly on the gate voltage, perhaps when filling the dot. 

\section{Interaction with static dots}
At NPL, Fletcher \etal \cite{Fletcher2003-2618} measured a channel that contained an unintentional quantum dot, which exhibited Coulomb-blockade (CB) oscillations. They found that the SAW-driven current plateaux observed when the channel was pinched off were correlated with the CB oscillations they observed when the channel was open. This led them to suggest that SAW-driven quantised current might \textit{always} require a pair of potential barriers in the channel (e.g.\ from impurities). The mechanism proposed is similar to the `turnstile' pump originally demonstrated by Kouwenhoven \cite{KouwenhovenPump1991}, which only worked at much lower frequencies (up to 10\,MHz). In such a pump two barriers go up and down in antiphase, letting an electron in and then out of the dot. For the SAW to modulate the static barriers in such a way would require them to be about half a SAW wavelength apart. The overall gate and power dependences were, however, different from those seen in normal channels, in that a criss-cross pattern of current steps emerged rather than a smooth evolution, with a plateau missed out on some gate-voltage sweeps. It seems unlikely, therefore, that this mechanism is really the cause of quantised current in channels without any apparent impurities or Coulomb-blockade oscillations. Indeed, Talyanskii wrote a detailed explanation of how the usual plateaux did not exhibit the claimed dot-like behaviour \cite{TalyanskiiComment2006}.

Ebbecke \etal then deliberately made a quantum dot with barriers a half wavelength ($0.5\,\mu$m) apart \cite{Ebbecke2004-2595} and saw a long $ef$ plateau with fairly good accuracy ($10^{-3}$), and it seemed that this might be promising as a current standard. Later, an unintentional impurity gave rise to a second quantum dot in one of the barriers, and a quantised plateau at very low SAW power \cite{Ebbecke2005-2810}. This double dot prompted the authors to propose entangling the spins of two electrons in a parallel pair of dots and emitting them both into a channel using a SAW, as a source of entangled electrons.

\begin{figure}[tb]%
	\includegraphics*[width=\linewidth]{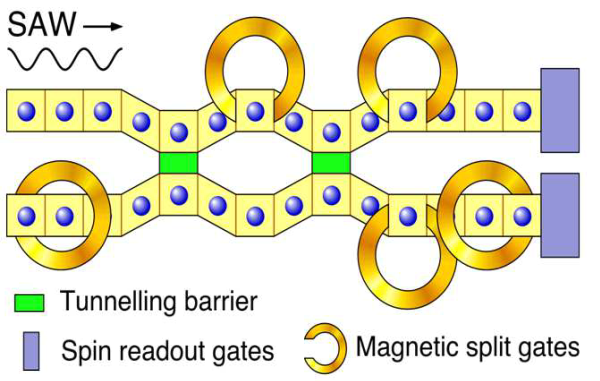}
	\caption{%
		Design for a C-NOT gate in the Barnes \etal proposal \cite{Barnes2000-2213} for a SAW-driven quantum processor. Yellow lines indicate channels containing single electrons (blue discs). A SAW is incident from the left. Horse-shoe shapes represent magnetic split gates and green rectangles represent tunnel barriers that produce `root of swap' operations. Spin readout or other operations occur in the purple rectangles at the right. Copyright 2000 by C.\,H.\,W.\,Barnes.}
	\label{SAWQCproposalb}
\end{figure}

Having concluded that a current standard was out of reach with current SAW devices, attention turned to the uses to which moving quantum dots could be put. In 2000, Barnes \etal \cite{Barnes2000-2213,Barnes2003-2587,Furuta2004-2654} proposed a quantum computer made from a set of electron-spin qubits driven along parallel channels by a SAW, past various structures to carry out one- and two-qubit operations. A number of the enabling technologies for such a computer have been developed since then, together with ways of integrating them with more conventional qubits based on static quantum dots. Firstly, Schneble \etal \cite{Schneble2006-2916} made a static (gate-defined) quantum dot and looked at how the Coulomb-blockade peaks broadened as the sample temperature was increased by application of a radio-frequency signal to a transducer. There was significant heating for powers above $\sim -15$\,dBm, even for frequencies off the SAW resonance so that there was no acoustic wave. Pulsing the SAW so that it was only on for a small fraction (0.01--0.03) of the time was enough to make the heating negligible at 300\,mK, even on the SAW resonance. Utko \etal saw similar heating using a 2DEG resistance as a thermometer and showed that the acoustic wave contributed only about half as much heating as the input radio-frequency signal itself \cite{UtkoHeating2006}.

\begin{figure}[tb]%
	\includegraphics*[width=\linewidth]{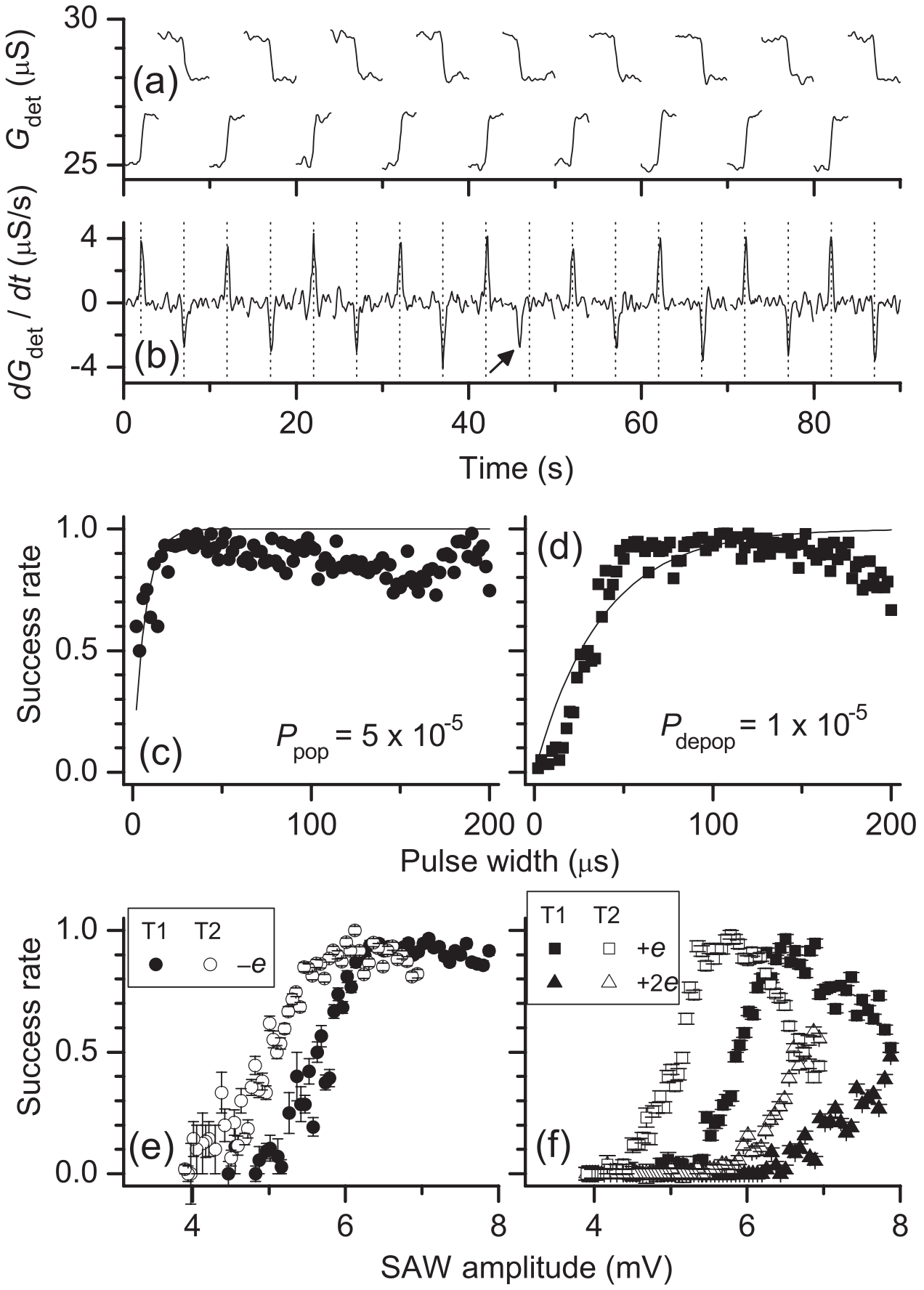}
	\caption{%
		Measurements of the effect of SAW pulses on a quantum dot out of equilibrium (from \cite{Kataoka2007-2892}). (a) Each trace shows the detector channel conductance $G_{\rm det}$ monitored before and after a short SAW pulse, which immediately causes a population or depopulation transition. SAW pulses are repeated every 5\,s, while the plunger gate is switched between the read and write positions in between pulses. (b) The derivative of the detector conductance in (a). The vertical dotted lines are plotted every 5\,s. The arrow marks a switching event that occurred prior to the SAW pulse. (c) and (d) The success rates of population and depopulation, respectively, as a function of pulse width. The solid lines are fits described in the main text. (e) and (f) The SAW-amplitude dependence of the success rates for population ($-e$) and depopulation ($+e$ and $+2e$), respectively, for the SAW transducers T1 and T2 at opposite ends of the sample. \href{http://dx.doi.org/10.1103/PhysRevLett.98.046801}{Reproduced with permission from Ref. \cite{Kataoka2007-2892}, copyright 2009 by the American Physical Society}.}
	\label{Kataoka2007stats}
\end{figure}

Schneble \etal next used their quantum dot with a nearby narrow 1D channel as a detector \cite{Field1993-133} of Coulomb charging in the dot \cite{Schneble2007-2915}. They pinched off the gates separating the dot from the source and drain leads, such that the tunnelling time for electrons trapped out of equilibrium in the dot became many seconds. When such an excess electron left the dot there was a sudden step in the 1D channel's conductance. Applying a SAW pulse made it much more likely that the excess electron would leave the dot immediately, by modulating the barriers and the potential of the dot itself. Similarly, an electron could enter the dot to fill an empty state below the fermi energy. Kataoka \etal used this same device to study the detailed statistics of `writing' (population) and `reading' (depopulation) of the dot out of equilibrium \cite{Kataoka2007-2892}, see Fig.\ \ref{Kataoka2007stats}. The dot could be tuned into readiness for either a read or a write operation just using a gate, offering the prospect of addressing a whole array of static dots individually by using a gate to tune a dot into the read or write mode. The pulse-length dependence was investigated in more detail \cite{Kataoka2008-3004}, showing that a pulse of just 20\,ns could depopulate the dot by one electron. Around the same time, in carbon nanotubes containing chains of disorder-induced quantum dots, Buitelaar \etal showed that a SAW could redistribute electrons between the dots \cite{Buitelaar2008-3073}.

\begin{figure}[tb]%
	\includegraphics*[width=\linewidth]{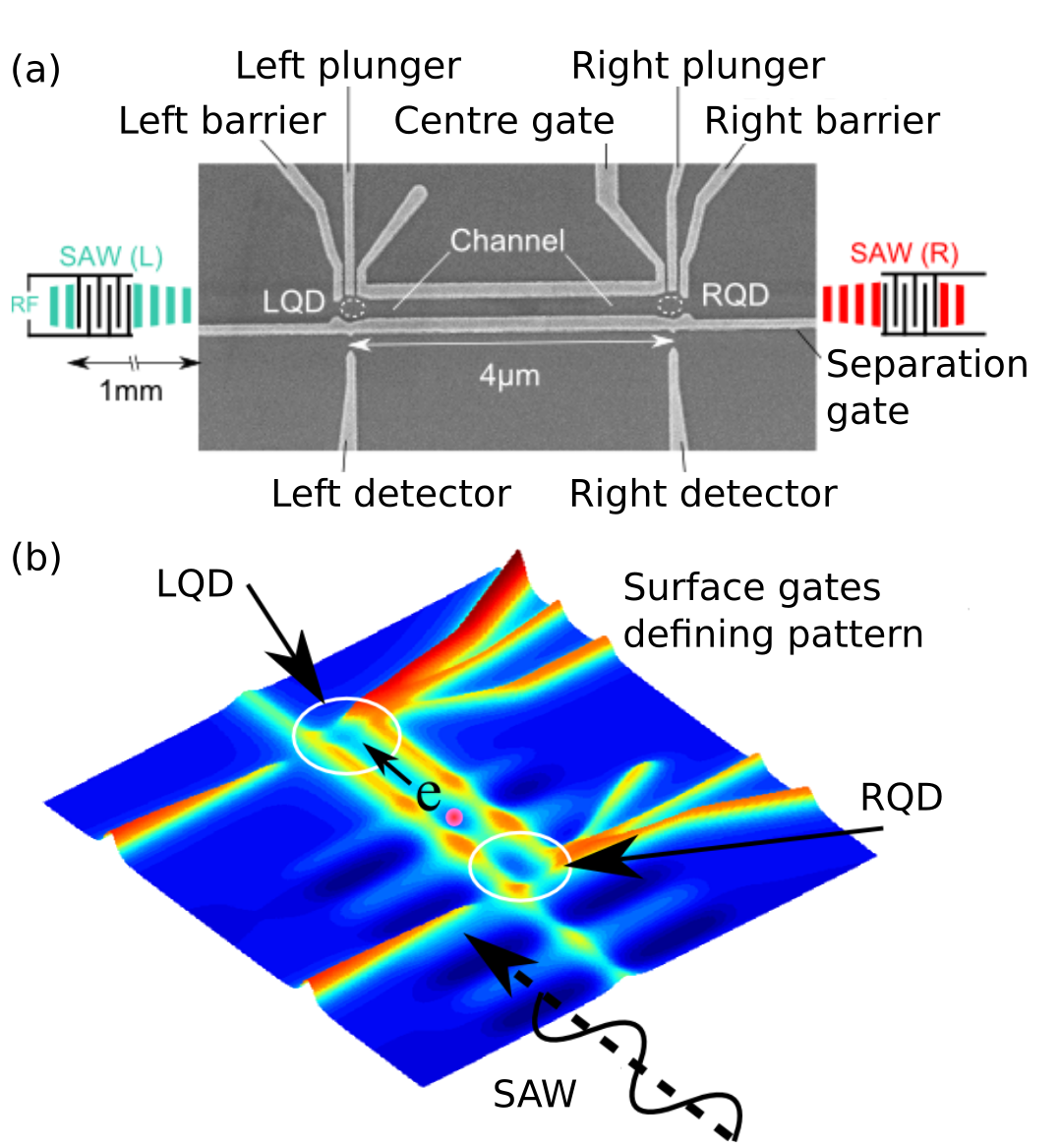}
	\caption{%
		(a) Electron micrograph of the dot--channel--dot device measured in \cite{McNeil2011-3370}; LQD and RQD are the left and right quantum dots, respectively, and all the gates, which appear light grey, are labelled with their names. (b) Intensity map of the calculated potential. As indicated, a SAW pulse arrives at RQD, taking the electron from there along the empty channel to the other, empty, dot, LQD. The detector constrictions are at the bottom left.}
	\label{QD2QDdiag}
\end{figure}

\begin{figure*}[tb]%
	\centerline{\includegraphics*[width=0.8\linewidth]{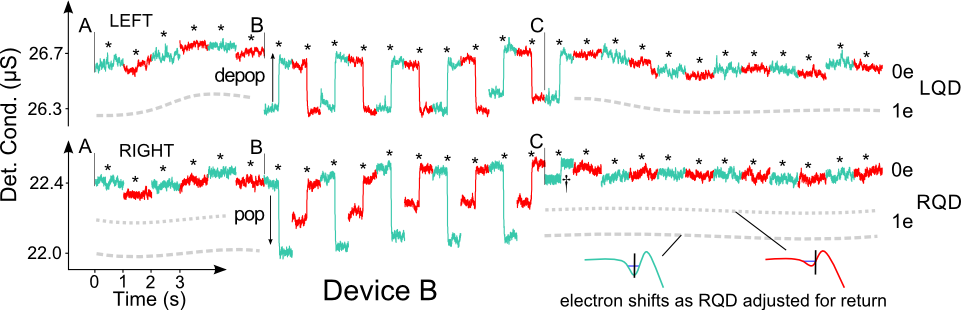}}
	\caption{%
		Single-electron `rally' in the dot--channel--dot device, reproduced with permission from \cite{McNeil2011-3370}: the two rows of data show the signal from left and right detectors, with a low value indicating the presence of an electron in the nearby dot. The quantum dots and the channel are initialized to be empty before time A. Between times A and B, a series of control pulses is used to verify that system is empty. At time B an electron is loaded into LQD. Between times B and C, there is two-way transfer of a single electron between the quantum dots. At time C, the electron is removed from the system using a clearing pulse. The SAW pulse duration is 300\,ns. The time between traces is not plotted.}
	\label{SAWQD2QDFig1d}
\end{figure*}

\section{Transfer of electrons between static dots}

Building on the earlier work with static dots, described in the previous section, in 2011 McNeil \etal in Cambridge added a channel as one exit from the static dot, and another dot at the other end of this channel, $4\,\mu$m away, as shown in Fig.\ \ref{QD2QDdiag} \cite{McNeil2011-3370}. Each dot had a 1D `detector' constriction beside it, so the charge state of each dot could be read out, and this showed that it was possible to depopulate each dot completely and to load an electron into it far above the Fermi energy, and to hold it there for many minutes by making the tunnel barriers very high. Likewise, the dot could instead remain empty. The channel was depleted completely by long side gates. Considerable optimisation was necessary to ensure that no charge became stuck in the channel or beside the dots, by bowing the centre gate inwards slightly towards the middle. At one end of the device, a transducer sent a SAW pulse that took an electron from the closest dot, along the channel, and deposited it in the other dot, see Fig.\ \ref{SAWQD2QDFig1d}. A transducer at the other end was then used to send a SAW pulse in the opposite direction, taking the same electron back along the channel to the first dot. This process could be repeated many times, with only occasional errors, usually when another electron entered the channel, presumably dragged in by the SAW from one of the leads. Such transfer of single-electron spin qubits could be very useful in a quantum computer consisting of static quantum dots, as it is currently very hard to move spin qubits over significant distances (many microns) to entangle them with other qubits, for example \cite{KontosComment2011}. 

Simultaneously, Hermelin \etal in Grenoble published very similar work \cite{Hermelin2013}, with the same geometry and scale. They had no working second transducer so could only send electrons in one direction, but they did have a gate at the exit from the first dot to which could be applied a 1\,ns pulse to let the electron out of the dot only within this pulse. Since then, the Grenoble group have investigated electron injection into their channel \cite{BertrandInj2016} and, most importantly, Bertrand \etal have used double dots at each end of a channel to prepare the spins of one or two electrons, transfer the electron(s) to the other end of the channel, and then measure the spin state in the other dots \cite{BertrandFast2016}. The classical fidelity of the whole spin-transfer procedure reaches 65\% and seems to be limited by depolarisation of the spins in the static dots before and after the SAW-driven transfer along the channel. This rapid spin transfer occurs in a much shorter time than the spin coherence time and so is suited to the time-scales required for quantum processing.

Unfortunately, GaAs systems have non-zero nuclear spin, which dephase the electron spins. Spin-echo techniques can be used to greatly reduce such dephasing \cite{BluhmDephasing2011} but only in static, not moving, dots, so there will always be some loss of fidelity during such a transfer, even though `motional narrowing' (averaging over many more nuclear spins) works in one's favour to reduce the dephasing \cite{Stotz2005,McNeil2011-3370}. Spin-orbit interactions produce an effective magnetic field that also causes spin precession. Stotz \etal \cite{Stotz2005,Stotz2008} made travelling arrays of quantum dots (containing 15--100 electrons each) at the intersecting minima of two SAWs propagating at right angles, and found spin-coherence times in excess of 10\,ns and a spin-coherence length over 100\,$\mu$m. Couto \etal \cite{Couto2007}, by investigating coherent manipulation of the spins of a collection of photoexcited electrons and holes transported in SAW minima and maxima, showed how the spin lifetime could be extended by suitable choice of wafer orientation and SAW direction. Huang and Hu later calculated that spin-orbit interaction with a disorder potential in the channel is a significant, but not necessarily dominant, source of spin dephasing \cite{Huang2013}. It would be advantageous to use silicon instead of GaAs as the spin lifetimes are very long there, but this requires an extra piezoelectric layer, such as ZnO \cite{Couto2007,Pedros2011-3371,Barros2012,Bukukkose2012}, to provide the SAW potential, as silicon itself is not piezoelectric, and the methods of defining dots in Si (e.g.\ multiple layers of gates) may screen the SAW so that electrons cannot be extracted from the dots. Manipulation of the spin qubits while moving along the channel, as proposed by Barnes \etal \cite{Barnes2000-2213}, may be achievable using nanomagnets on either side of the channel. For this, McNeil \etal have developed and tested a variety of nanomagnet designs to provide fields pointing in arbitrary directions but set up by a uniform external field such as those available in many cryostats \cite{McNeil2010-3126}. 

\section{Non-adiabatic excitation in moving dots}
One element of the quantum computer proposed by Barnes \etal \cite{Barnes2000-2213} is a tunnel barrier between two SAW-driven dots (see Fig.\ \ref{SAWQCproposalb}). Single electrons in each dot would either swap over or not, leading their wave-functions to become entangled (double occupation of one dot would be energetically unfavourable due to their mutual Coulomb repulsion). Kataoka \etal made a device containing two such parallel channels and tried to see a suppression of tunnelling across the tunnel barrier when the other dot was occupied compared with when it was empty \cite{Kataoka2006-2774}. However, no such effect was seen, and it is likely that the tunnel barrier between the two parallel channels was not necessarily the same height all along its length, and an electron could be tipped over the top of it with excess energy, enough to overcome the Coulomb repulsion.

\begin{figure}[tb]%
	\includegraphics*[width=\linewidth]{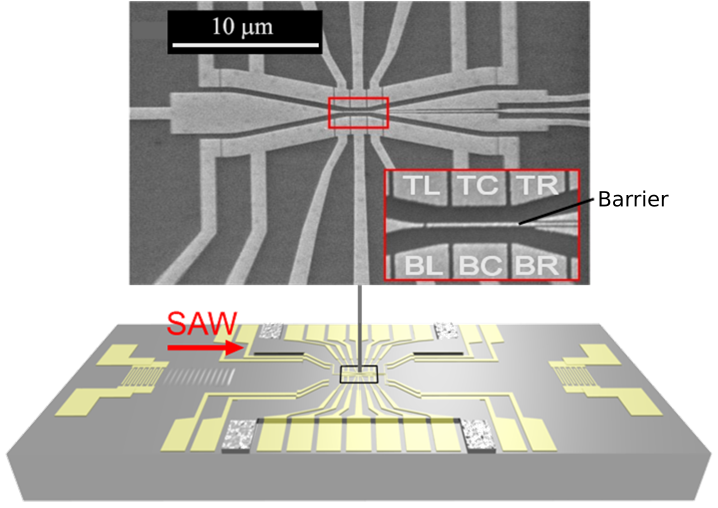}
	\caption{%
		Illustration of the two-channel device, including the SAW transducers at the left and right ends of the chip (bottom) and electron micrograph showing the surface gates (top). The inset shows the central $3.5\,\mu{\rm m} \times 2.5\,\mu$m of the device. The narrow gaps between gates are completely pinched off, allowing separate control of each section of the channels. \href{http://link.aps.org/doi/10.1103/PhysRevLett.102.156801}{Reproduced with permission from Ref.\ \cite{Kataoka2009-3029}, copyright 2009 by the American Physical Society}.}
	\label{SAW1TunnelBarrier}
\end{figure}

Astley and Kataoka \etal \cite{Astley2007-2959,Kataoka2009-3029} next made a much more sophisticated pair of channels with 18 gates around them to give great control over each region, as shown in Fig.\ \ref{SAW1TunnelBarrier}. Astley \etal studied energy-dependent tunnelling from a dot in one channel to a much wider region on the other side of the tunnel barrier, which was therefore also lower in energy \cite{Astley2007-2959}. When injecting 1, 2 or 3 electrons in each SAW minimum (with reasonable accuracy), they measured the loss of current from the channel and hence the probability of tunnelling across the barrier in each case, as a function of barrier height. Tunnelling could only occur during the $\sim 600$\,ps during which the SAW-driven dot was passing beside the $2\,\mu$m-long barrier. The energy differences between the dots containing 1, 2 and 3 electrons were estimated, using a model potential and rate equations, to be a few meV or more, which is mainly the charging energy of adding an extra electron to the dot. 

\begin{figure}[tb]%
	\includegraphics*[width=\linewidth]{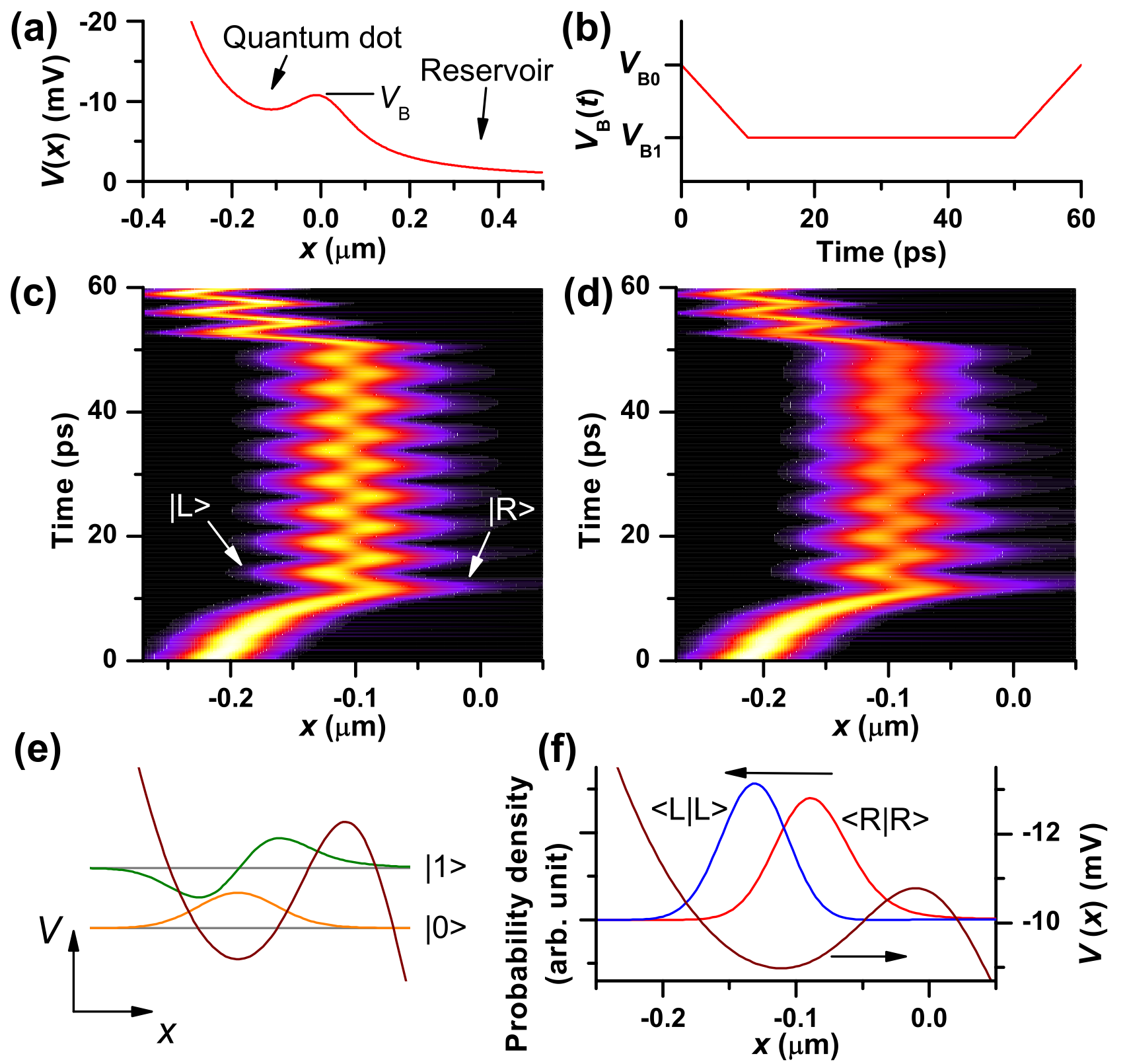}
	\caption{%
		Time-dependent 1D model demonstrating the oscillating tunnel rate in the two-channel device (from \cite{Kataoka2009-3029}). (a) Calculated potential $V$ across the tunnel barrier. (b) Barrier-gate potential $V_{\rm B}$ \textit{vs} time. (c) and (d) Time evolution of the probability density of the wave function plotted as a colour scale (a brighter colour shows a higher value) for the cases of a high tunnel barrier (c) and a low tunnel barrier (d). (f) Probability densities at the left and right extreme positions of the wave function, $|$L$\rangle$ (blue) and $|$R$\rangle$ (red), respectively, which can be approximated as linear combinations of the ground state $|$0$\rangle$ and first excited state $|$1$\rangle$ shown in (e). The potential profile is plotted in brown. \href{http://link.aps.org/doi/10.1103/PhysRevLett.102.156801}{Reproduced with permission from Ref.\ \cite{Kataoka2009-3029}, copyright 2009 by the American Physical Society}.}
	\label{Kataoka2009model}
\end{figure}

Using the same device configuration, Kataoka \etal noticed very small but reproducible oscillations in the tunnelling probability as a function of the gate voltages around the barrier. After carefully ruling out explanations related, for example, to weak points due to impurities in the barrier \cite{Kataoka2008-3001}, Thorn \cite{ThornThesis2009} solved the 1D time-dependent Schr{\"o}dinger equation for transverse slices of the realistic potential profile in the rest frame of the SAW. He saw similar oscillations, with the right gate-voltage dependences  \cite{Kataoka2009-3029}. What happens is that the electron is initially travelling along a channel at an angle to the SAW propagation direction (top left of the inset in Fig.\ \ref{SAW1TunnelBarrier}), but then it changes direction slightly to travel along the side of the barrier, and the dot opens up as the side-wall potential drops at the tunnel barrier. The change in potential at this point occurs within about 40\,ps, which is too fast for the ground-state wave function to evolve smoothly into the new ground state. This non-adiabatic change leaves the electron in a superposition of ground and excited states of the new dot potential. Those with the highest energy can get over or through the tunnel barrier easily, leaving effectively just the ground and first excited states. The phase of each state evolves differently in time and so the sum of these is a wave packet that oscillates from side to side in the channel as it travels along beside the barrier, with frequency equal to the energy difference between the states divided by Planck's constant $h$, see Fig.\ \ref{Kataoka2009model}.

While moving along the tunnel barrier the electron has a number of opportunities to tunnel out when the oscillation brings it close to the barrier. The total probability $P_{\rm tot}$ of leaving is the combination of these individual probabilities. As a gate voltage is made more negative, the confining potential in the channel becomes stronger, increasing the oscillation frequency and hence reducing the time between `bounces' off the barrier. Initially the time spent near the barrier during each bounce drops, reducing  $P_{\rm tot}$, but as the gate voltage continues to change another bounce can be fitted in during passage along the barrier, so $P_{\rm tot}$ increases again. These oscillating changes in the tunnelling current are very small ($\sim 1\%$), meaning that there is little effect on the electron wave function in the channel, i.e.\ it is not a precise measurement of the electron's position or momentum and so the wave function does not immediately collapse. This process is repeated on each cycle so successive electrons build up a detectable current. This technique of moving a particle past a change in static potential and/or making a `weak' measurement of its wave function could be very powerful for making and investigating non-adiabatic excitations on a time-scale considerably shorter than currently achievable using pulses applied externally to gates on the sample in a cryostat.

\section{Single-photon sources}
For quantum cryptography, a controlled source of single photons is required, with high repetition rate. 
Photoluminescence has already been extensively investigated and controlled using SAWs to separate photoexcited electrons and holes so that they do not immediately recombine or interact. In particular, Rocke \etal \cite{Rocke1997} used a laser pulse in one place to excite excitons that were split up and transported over a distance of at least 1\,mm before the SAW moved under a metal gate that screened the SAW and hence allowed recombination, with the accompanying emission of photons. The interaction of SAWs and light in semiconductor structures has been reviewed in detail by de Lima and Santos \cite{deLimaReview2005}.

Foden \cite{Foden2000-2253} proposed that a SAW could drag single electrons up a potential slope into a region of holes, where they would recombine to produce a stream of single photons, at GHz frequencies. The quantisation accuracy would not need to be high, as occasional missing photons would not be serious for cryptography. To fabricate such a device, nearby regions of electrons and holes are needed, which is impossible to achieve by conventional material growth by molecular-beam epitaxy (MBE). In 2004, Hosey \etal in Cambridge used a special MBE chamber with an \textit{in-situ} 30\,keV focussed-ion beam to pattern different regions with Be (\textit{p}-type) or Si (\textit{n}-type) dopants \cite{Hosey2004-2523}. A SAW did indeed drag electrons across to the holes, where they recombined to produce light, but this MBE machine could only produce wafers 1\,cm on a side, so it was not practical to make many samples in this way. Cecchini \etal in Pisa developed lateral \textit{p-n} junctions by etching away a Be acceptor layer in some places and depositing an n-type contact there \cite{Cecchini2005-2769}. A SAW caused electroluminescence oscillations in time with the same period as the SAW, though the forward bias was so large that the SAW might just have been helping the bias by suppressing the potential barrier, rather than dragging electrons in SAW minima. Smith \etal doped an InSb/Al$_x$In$_{1-x}$Sb heterostructure with layers of p and n impurities, and developed a bevel-etching technique to remove the p-type layer at one end (making it n-type) to form a lateral n-i-p junction \cite{SmithNash2006} with a graded potential that rose slowly enough for a SAW to transport electrons across the junction into the p-type region \cite{SmithNash2007}. 

In 2007, Gell \etal \cite{Gell2007-2958} etched a (311)A crystal facet on a line across in an MBE-grown GaAs wafer and grew more GaAs and AlGaAs layers, with Si doping. On the facet, the doping was p-type, and on the usual (100) surface it was n-type. When a SAW was applied travelling towards the facet, photons were produced at a particular part of each SAW cycle, presumably corresponding to packets of electrons being taken in SAW minima across the intrinsic region into the facet. This processing technique was also very time-consuming and so it may be more satisfactory to induce both electrons and holes in an undoped wafer. Such a lateral \textit{p-n} junction was fabricated in Pisa a few years later: De Simoni \etal showed that they could use a SAW to pump a stream of electrons into the sea of holes, and they observed a corresponding flux of photons as electrons and holes recombined \cite{DeSimoniAPL2009}. In Cambridge, induced samples have now shown quantised current when pumping with a SAW from one region of electrons to another \cite{ChungHou2016}, but so far no quantisation when pumping from electrons to holes.

In principle, such induced samples should offer better current quantisation if suitable channel geometries can be developed, as impurities are almost completely absent in high-quality undoped material, so there are no bumpy potentials or traps along depleted channels and no random-telegraph switching noise. Talyanskii showed how a narrow gate just downstream of a parallel inducing gate could be used to help a SAW extract electrons from the induced region into a broad intrinsic region from which they were funnelled into a narrow channel \cite{Talyanskii2006-2756}. In that channel was a centre gate to split the beam into two parts, so that electrons in some SAW minima could be diverted into one of two channels, forming a Y-branch switch. Later, they showed that pulses lasting just 100\,ns or less applied to the extraction gate and a side gate could launch packets of electrons and switch them between branches \cite{TalyanskiiFidelity2008}. Talyanskii \etal also proposed a means of detecting photons using SAWs, in a similar way \cite{TalyanskiiSST2007}.

B\"{o}defeld \etal \cite{Bodefeld2006} made a photon source using dots grown in a quantum well by self-assembly of a strained material (InP) and later Lazi\'{c} \etal developed this technique by patterning an array of triangular dots and then growing a quantum well on top by MBE \cite{Lazic2012}. They showed antibunching of photon emission, a precursor to full suppression of multiple photon emission.

\section{Future prospects}
Single-electron quantum dots driven along channels by surface acoustic waves are unlikely to form the basis of a current standard unless some new device design is found. However, even with an accuracy of 1 in $10^3$ or $10^4$, other applications are possible. The ability to move spin qubits between distant static dots could be important in quantum computing. Also, one can imagine a quantum `repeater' where a SAW picks up electron-hole pairs created by single photons hitting one region, brings them to static dots or channels for storage, processing, measurement  or entanglement (as just electrons or both electrons and holes), and then takes them further to emit them as photons again. In order to maximise the spin-coherence time, silicon-based devices (with their lack of nuclear spins) may be needed, though novel quantum-dot designs will be needed to make it easy for a SAW to extract electrons from them, without the gates screening the SAW.

\begin{acknowledgement}
I acknowledge with gratitude the work of many researchers in the Semiconductor Physics Group at the Cavendish Laboratory in Cambridge over the past two decades, particularly Dr Valery Talyanskii, Prof Michael Pepper, Prof Charles Smith, Prof Crispin Barnes, Julian Shilton, John Cunningham, Andy Robinson, Fran\c{c}ois Sfigakis, Peter Leek, Dr Masaya Kataoka, Dr Mark Graham, Mike Astley, Jeff Schneble, Jenny Gell, Adam Thorn, Dr Mark Buitelaar, Dr Stuart Holmes, Rob McNeil, Matthew Benesh, Seok-Kyun Son, Yousun Chung, and Hangtian Hou, together with fruitful collaborations with NPL and Toshiba Cambridge Research Europe Ltd.
\end{acknowledgement}

%
\bibliography{saw_bib_with_hyperlinks}
\bibliographystyle{pss}
%


%
%
%
%
%
%


\end{document}